\documentstyle[osa]{revtex}

\begin{document}


\title{Coexistence of thermal noise and squeezing in the 
intensity fluctuations of small laser diodes}

\author{Holger F. Hofmann}
\address{Department of Physics, Faculty of Science, University of Tokyo\\
7-3-1 Hongo, Bunkyo-ku, Tokyo113-0033, Japan}

\author{Ortwin Hess}
\address{Theoretical Quantum Electronics, Institute of Technical Physics,
Deutsches Zentrum f\"ur Luft- und Raumfahrt,
\\ Pfaffenwaldring 38-40, D-70569 Stuttgart, Germany}


\date{\today}

\maketitle

\begin{abstract}
The intensity fluctuations of laser light are derived from photon number
rate equations. In the limit of short times, the photon statistics for
small laser devices such as typical semiconductor laser diodes show thermal
characteristics even above threshold. In the limit of long time averages 
represented by the low frequency component of the noise, the same devices 
exhibit squeezing. It is shown that squeezing and thermal noise can coexist 
in the multi-mode output field of laser diodes. This result implies that 
the squeezed light generated by regularly pumped semiconductor laser diodes
is qualitatively different from single mode squeezed light. In particular,
no entanglement between photons can be generated using this type of collective
multi-mode squeezing.
\end{abstract}

\section{Introduction}

In the early days of laser physics, one of the most fundamental properties
attributed to lasers was the reduction of photon number fluctuations
below thermal levels at threshold \cite{Hak64,Scu67,Lax69,Ris70,Hak72}. 
More recently, the possibility to
reduce the intensity noise of semiconductor laser diodes even below the
shot noise limit has once again drawn attention to the photon number 
statistics of light emitted by laser devices 
\cite{Yam86,Mch87,Mch88,Mch89,Ino92,Fre93,Kil96}. 
However, in the case of semiconductor lasers, the microscopic
laser dynamics depends sensitively on the charge carrier
densities which provide the optical gain. Quantum theories of laser
light often fail to include this dynamical degree of freedom. In particular, 
the theories which show the reduction of photon number fluctuations in the
laser cavity below thermal noise usually rely on the adiabatic elimination
of the carrier dynamics \cite{Ken89,Wal94}. 
In the following, 
it is shown that, without this elimination of the carrier relaxation 
dynamics, small semiconductor lasers may exhibit thermal noise even far above
threshold. In particular, the photon number fluctuation inside the cavity
may even be thermal at quantum efficiencies grater than 50 \%, allowing a
suppression of the low frequency intensity fluctuations below the shot noise
limit. 
This coexistence of thermal noise and squeezing can be derived theoretically 
from the same photon number rate equations as the thermal noise. It is 
therefore necessary to distinguish the photon number 
statistics at short times from the time integrated statistics of the low
frequency noise component. Specifically, the low frequency noise is not
associated with any well defined mode, but represents a collective property
of many modes
due to the limited temporal coherence of the emitted light. This implies 
that even an ideal single mode laser exhibits the anti-correlated
intensity statistics observed in multi-mode lasers \cite{Ino92,Mar95,Esc96}. 
Even the squeezing observed in a perfect single mode laser can therefore 
include a large amount of thermal noise and does not usually correspond to
the generation of a minimum uncertainty state.

The paper is organized as follows:
In section \ref{sec:rate}, the rate equations are formulated. Within this
framework we present our definition of the {\em laser threshold}. 
In the subsequent section \ref{sec:pnf}, the photon number fluctuations 
are determined which allow the definition of a {\em noise threshold}. 
In section \ref{sec:lfn}, the low frequency limit of intensity noise 
is derived and the {\em squeezing threshold} is defined. 
In section \ref{sec:sum}, the results are summarized 
and the conditions for a coexistence of squeezing and thermal noise is 
obtained.
In section \ref{sec:col}, the nature of collective multi-mode squeezing is 
discussed and limitations of quantum optical applications are pointed 
out. 


\section{Rate equations for the energy flow in lasers}
\label{sec:rate}

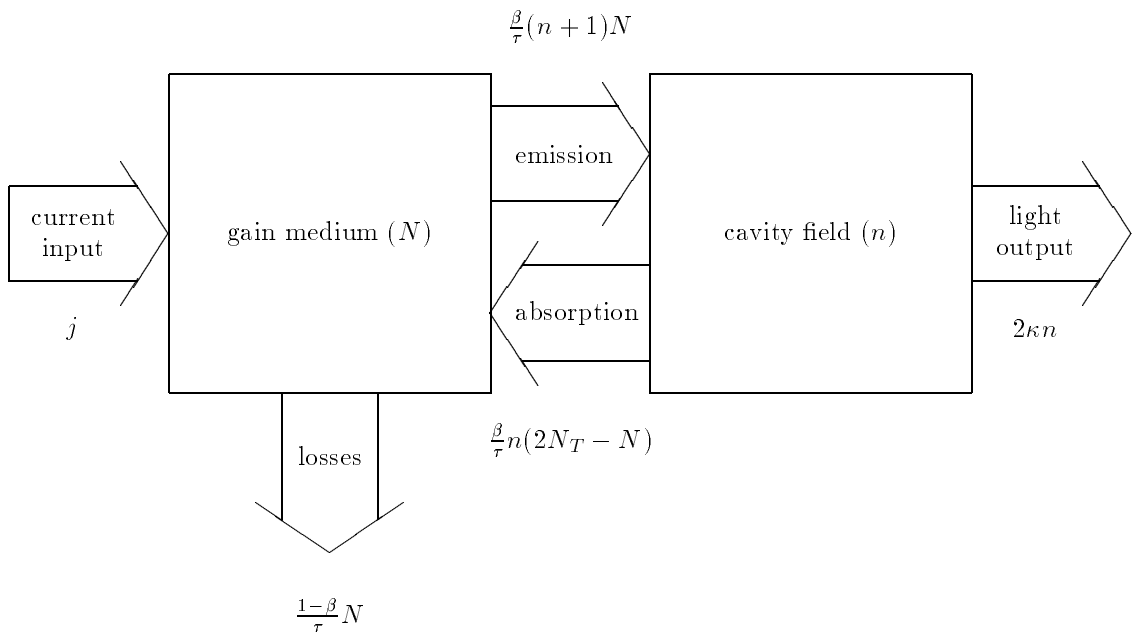
\begin{figure}
\caption{Energy flow diagram of a single mode laser}
\label{energyflow}
\end{figure}

A laser device converts the energy injected into the gain medium
into laser light by means of light field amplification inside the
laser cavity. 
Since the energy in the gain medium and the energy
of the light field are both quantized, the energy flow cannot be 
entirely smooth and continuous. Instead, it is described by transition
rates. Figure \ref{energyflow} illustrates the transition rates
between the two energy reservoirs and the environment. Note that
the rates given pertain to a single mode laser under the assumption of
a linear carrier density 
dependence for gain and spontaneous emission.
The laser device is thus characterized by 
the spontaneous emission factor $\beta$, 
the excitation lifetime $\tau$, 
the excitation number at transparency $N_T$ and 
the cavity loss rate $\kappa$. 
The rate equations for the excitation number $N$ and the cavity photon 
number $n$ at an injection rate $j$ may be formulated 
according to figure \ref{energyflow} as
\begin{eqnarray}
\label{eq:energyflow}
\frac{d}{dt} N &=& j-\frac{1}{\tau} N - 2\frac{\beta}{\tau}(N-N_T)n
+ q_N(t)
\nonumber \\
\frac{d}{dt} n &=& 2\left(\frac{\beta}{\tau}(N-N_T)-\kappa\right) n
     + \frac{\beta}{\tau}N + q_n(t),
\end{eqnarray}
where $q_N(t)$ and $q_n(t)$ represent the shot noise terms corresponding to
the respective transitions into and out of the gain medium excitation $N$
and the cavity photon number $n$, respectively. 
Note that equation (\ref{eq:energyflow}) corresponds to the rate equation
of Lax and Louisell \cite{Lax69}. Although a precise quantum mechanical 
treatment would require the solution of a master equation for both the 
light field and the gain medium, it is possible to reduce this equation 
to the diagonal elements in photon and excitation number. The statistical 
equations then correspond to those of a classical particle flow, allowing 
a formulation of both rate equations and shot noise terms. 

The noise terms are zero on average. However, their fluctuations and 
correlations are given by the transition rates associated with the 
respective energy reservoir,
\begin{eqnarray}
\label{eq:qcorr}
\langle q_N(t)q_N(t+\Delta t)\rangle &=& \left( \sigma j + 
\frac{\beta}{\tau}(n+1)N
+ 2\frac{\beta}{\tau}N_T n +\frac{\beta}{\tau} N \right)\;\delta(\Delta t)
\nonumber \\ 
\langle q_n(t)q_n(t+\Delta t)\rangle &=& \left( 2\kappa n
+ 2\frac{\beta}{\tau}N_T n +\frac{\beta}{\tau} N \right)\;\delta(\Delta t)
\nonumber \\
\langle q_N(t)q_n(t+\Delta t)\rangle &=& - 
\left(2\frac{\beta}{\tau}N_T n +\frac{\beta}{\tau} N \right)\;\delta(\Delta t).
\end{eqnarray}
Pump noise suppression is described by the pump noise factor $\sigma$.
For the purpose of describing the possibility of squeezing, only the 
ideal case of $\sigma=0$ will be considered.
Moreover, it should be noted that the actual output intensity $I(t)$
of the laser device is a fluctuating quantity given by
\begin{equation}
\label{eq:intense}
I(t)= 2\kappa n + q_I(t),
\end{equation}
where the quantum noise statistics of $q_I(t)$ read
\begin{eqnarray}
\label{eq:output}
\langle q_I(t)q_I(t+\Delta t)\rangle &=&  2\kappa n \;\delta(\Delta t)
\nonumber \\[0.1cm]
\langle q_n(t)q_I(t+\Delta t)\rangle &=& - 2\kappa n \;\delta(\Delta t)
\nonumber \\[0.1cm]
\langle q_N(t)q_I(t+\Delta t)\rangle &=& 0.
\end{eqnarray}
Within the framework of our general laser model,
equations (\ref{eq:energyflow}) to (\ref{eq:output}) provide a complete
description of the fluctuating laser intensity $I(t)$ for all time-scales
and frequencies. The characteristics of a specific device are determined 
by only four device parameters, i.e.~the spontaneous emission factor
$\beta$, the excitation lifetime $\tau$, the excitation number at 
transparency $N_T$ and the cavity loss rate $\kappa$. In general, these
parameters are given by the gain medium and the cavity design. For 
semiconductor lasers with an active volume of $V$, typical material 
properties are 
\cite{Ebe92}
\begin{eqnarray} 
\label{eq:matpar}
\beta V         &\approx& 10^{-14}\mbox{cm}^{3} 
\nonumber \\
\frac{N_T}{V}   &\approx& 10^{18} \mbox{cm}^{-3} 
\nonumber \\
\tau            &\approx& 3 \times 10^{-9} \mbox{s}.  
\end{eqnarray}
The cavity loss rate $\kappa$ should be smaller than the maximal
gain in order to achieve lasing. For the parameters above, this 
corresponds to the condition that
\begin{equation}
\kappa < \frac{\beta N_T}{\tau} \approx 3.33 \times 10^{12} \mbox{s}^{-1}.
\end{equation}
Usually, the quality of a laser cavity will not be much higher than needed,
so the loss rate for semiconductor laser cavities is typically around
$10^{12} \mbox{s}^{-1}$. Consequently, the main device parameter for
semiconductor laser diodes is the size as given by $V$ or, alternatively, 
by the spontaneous emission factor $\beta$. Since the size of typical 
laser diodes is in the micrometer range, spontaneous emission factors for
standard diodes range from $\beta=10^{-3}$ for small vertical cavity surface 
emitting lasers to about $\beta=10^{-6}$ for edge emitters. Although larger
devices are possible, it becomes increasingly difficult to stabilize a 
single mode. Consequently, a single mode theory will tend to underestimate
the laser noise observed in large semiconductor laser diodes.  

The light-current characteristic of a laser device described by equation
(\ref{energyflow}) may be obtained from the time averaged energy flow
given by the stationary solution of the rate equations. The stationary 
carrier number average $\bar{N}$ and the stationary photon number average
$\bar{n}$ read
\begin{eqnarray}
\label{eq:statio}
\bar{N} &=& \frac{1 + \frac{1}{2 n_T}}{1+\frac{1}{2 \bar{n}}} N_T
\nonumber \\
\bar{n} &=& \frac{j-j_{th}}{4\kappa}-\frac{1}{4}
        +\sqrt{(\frac{j-j_{th}}{4\kappa}+\frac{1}{4})^2
               + \frac{j_{th}}{4\kappa}},
\end{eqnarray}
where the photon number at transparency $n_T$ and 
the threshold current $j_{th}$ are given by
\begin{eqnarray}
\label{eq:jtnt}
n_T &=& \frac{\beta N_T}{2\kappa\tau}
\nonumber \\
j_{th} &=& lim_{\bar{n}\to\infty} (1-\beta)\frac{\bar{N}}{\tau}
= 
2\kappa \frac{1-\beta}{\beta} \left(n_T+\frac{1}{2}\right).
\end{eqnarray}
The threshold current is defined by extrapolating the asymptotic linear 
increase of $\bar{n}(j)$ far above threshold to the threshold region. 
Since the excitation number $\bar{N}$ is pinned for $\bar{n}\to\infty$, 
the extrapolated threshold current is equal to the constant loss rate
far above threshold. 

In the following, the noise properties of the light field will be discussed
with respect to the light field intensity given in terms of the average
photon number $\bar{n}$. It is therefore  useful to define the photon number
at threshold by
\begin{equation}
\label{eq:nthdef}
n_{th} = \bar{n}(j_{th}) 
       = \sqrt{\frac{1}{16}+\frac{j_{th}}{4\kappa}} - \frac{1}{4}.
\end{equation}
Note that the order of magnitude for both threshold current $j_{th}$ and
threshold photon number $n_{th}$ is defined by the spontaneous emission factor
$\beta$. For $\beta\ll 1$ and $n_T=3/2$, the threshold current $j_{th}$ 
given by equation (\ref{eq:jtnt}) is
equal to $4\kappa/\beta$ and the photon number at threshold is $\beta^{-1/2}$.
In terms of electrical currents, $\kappa=10^{12}\mbox{s}^{-1}$ corresponds
to $1.6 \times 10^{-7} \mbox{A}$. Therefore, the approximate electrical
threshold current $\mbox{I}_{th}$
of a typical semiconductor laser diode is related to the spontaneous emission
factor $\beta$ by $2 \beta \mbox{I}_{th}\approx 10^{-6}\mbox{A}$. This formula
allows a simple quantitative estimate of the spontaneous emission factor from 
the threshold current. 
For example, a threshold current of 5 mA indicates a spontaneous emission 
factor of $\beta=10^{-4}$. The noise properties discussed in the following
can thus be related directly to the electrical threshold current observed
in the light-current characteristics of laser diodes.


\section{Photon number fluctuations}
\label{sec:pnf}

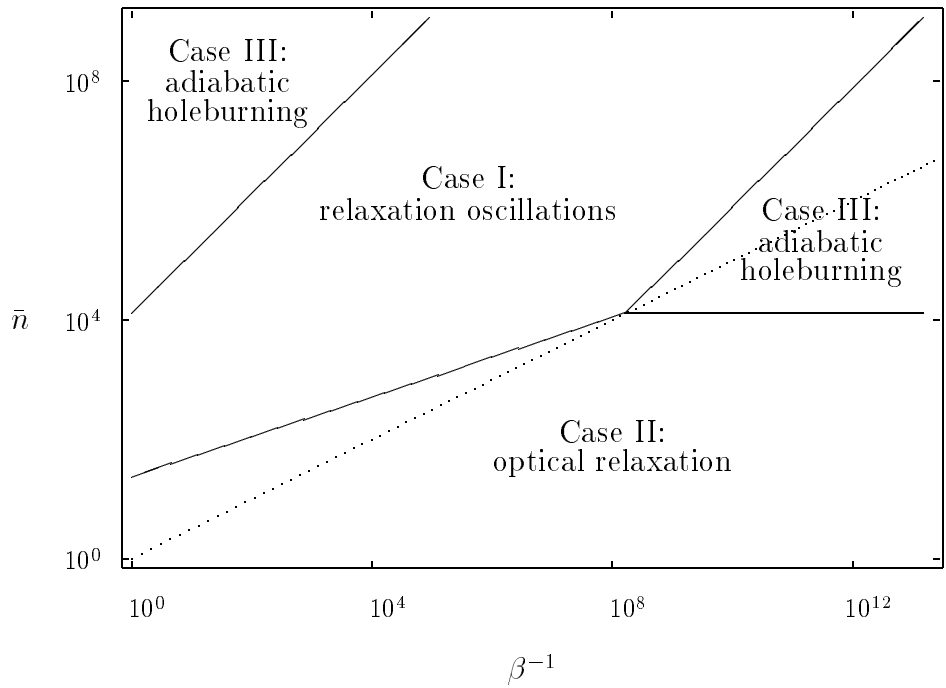
\begin{figure}
\caption{Dominant timescales of the fluctuation dynamics by 
average photon number $\bar{n}$ and spontaneous emission factor $\beta$
for $n_T=3/2$ and $3\kappa\tau=10^4$. The threshold
photon number $n_{th}$ is given by the dotted line.}
\label{cases}
\end{figure}

The noise characteristics of laser light can be investigated by solving
the linearized Langevin equations near the stationary solution. 
The linearization neglects the bilinear modification of the transition rate
given by $2 \beta \tau^{-1} \delta\!N \delta\!n$. This modification may become
relevant for large correlated fluctuations in both $n$ and $N$. However,
even for large $\delta\!n$, the smallness of $\delta\!N$ and the lack of
correlation between $\delta\!n$ and $\delta\!N$ usually justify the linear
approximation. 
The linearized
dynamics of the fluctuations $\delta\!N = N-\bar{N}$ and 
$\delta\!n = n-\bar{n}$ read 
\begin{eqnarray}
\label{eq:scalefluct}
\frac{d}{dt} \delta\!N 
       &=& -\Gamma_N \delta\!N - r\;\omega_R \delta\!n + q_N(t)
\nonumber \\ 
\frac{d}{dt} \delta\!n 
       &=& r^{-1}\;\omega_R \delta\!N -\gamma_n \delta\!n + q_n(t),
\end{eqnarray}
where the relevant time-scales are given by the electronic relaxation rate 
$\Gamma_N$, the optical relaxation rate $\gamma_n$, and 
the coupling frequency $\omega_R$. The hole-burning ratio $r$ scales the
interaction between photon fluctuations and excitation fluctuations.
The four parameters characterizing the fluctuation dynamics are functions
of the device properties and the average photon number $\bar{n}$, which read
\begin{eqnarray}
\label{eq:ratedef}
\Gamma_N &=& \frac{1}{\tau}(1+2\beta \bar{n}) 
\nonumber \\
\gamma_n &=& 2\kappa \frac{n_T+\frac{1}{2}}{\bar{n}+\frac{1}{2}} 
\nonumber \\
\omega_R &=& 2\kappa \sqrt{\beta \frac{\bar{n}-n_T}{\kappa\tau}}
\nonumber \\
r        &=& \sqrt{\frac{\kappa\tau}{\beta}
                   \frac{(\bar{n}-n_T)}{(\bar{n}+\frac{1}{2})^2}}.
\end{eqnarray}
The complete set of two time correlation functions describing the
temporal fluctuations may now be derived analytically by 
obtaining the response function of the linear dynamics and applying it
to the statistics of the noise input components $q_N(t)$ and $q_n(t)$.
However, it is usually possible to identify the major dynamical
processes observable in the fluctuation dynamics by concentrating only 
on the fastest time-scales. In particular, three regimes may be distinguished:
\newcounter{case}
\begin{list}{\Roman{case}.}{\usecounter{case}}
\item {\bf Relaxation oscillations}, $\omega_R\gg \Gamma_N+\gamma_n$ \\
  If the coupling frequency $\omega_R$ is much larger than the relaxation
  rates $\Gamma_N$ and $\gamma_n$, the fluctuation dynamics is described
  by relaxation oscillations with a frequency of $\omega_R$ and a relaxation
  rate of $(\Gamma_N+\gamma_n)/2$.

\item {\bf Optical relaxation}, $\gamma_n \gg \Gamma_N + \omega_R$ \\
  If the optical relaxation rate $\gamma_n$ is much larger than the 
  electronic relaxation rate $\Gamma_N$ and the coupling frequency 
  $\omega_R$, the excitation dynamics has no significant effect on 
  the fluctuation dynamics of the light field. 
  The fluctuation dynamics is then approximately described by thermal
  fluctuations with a coherence time of $\gamma_n^{-1}$. 
  
\item {\bf Adiabatic hole-burning}, $\Gamma_N \gg \gamma_n + \omega_R$ \\
  If the electronic relaxation rate $\Gamma_N$ is much larger than the
  optical relaxation rate $\gamma_n$ and the coupling frequency 
  $\omega_R$, the excitation number quickly relaxes to the stationary
  value defined by the much slower photon number fluctuations. This 
  stationary value of the excitation number acts back on the photon 
  number fluctuation through the coupling rate $\omega_R$, increasing 
  the relaxation rate in the light field by $\omega_R^2/\Gamma_N$. 
  The fluctuation dynamics is then given by exponentially damped fluctuations
  which are thermal for $\gamma_n\Gamma_N > \omega_R^2$ and become sub-thermal
  for $\gamma_n\Gamma_N < \omega_R^2$. In large lasers, this solution is 
  typically valid close to threshold.    
\end{list}
Figure \ref{cases} shows the operating regimes corresponding to the
three cases given above as a function of spontaneous emission factor
$\beta$ and average photon number $\bar{n}$.

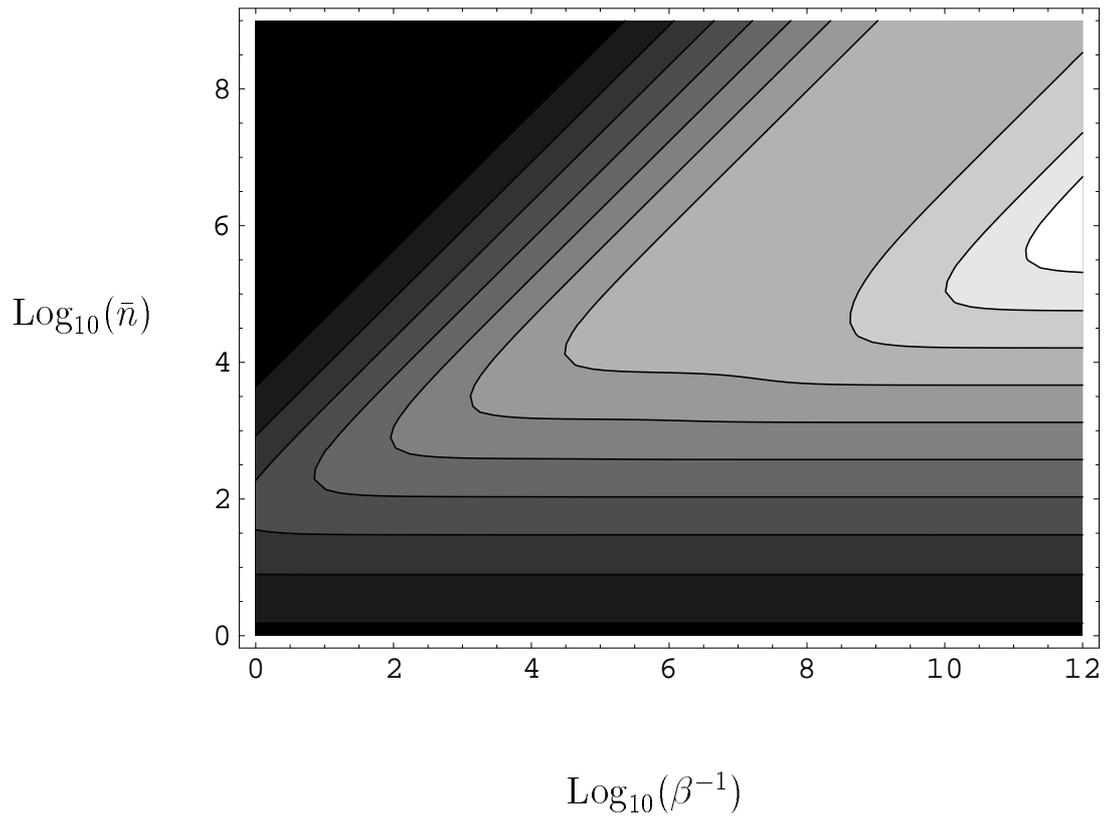
\begin{figure}
\caption{Contour plot of the photon number fluctuations as a function of
average photon number $\bar{n}$ and spontaneous emission factor $\beta$
for $n_T=3/2$ and $3\kappa\tau=10^4$. The contours correspond to constant
ratios of the photon number fluctuations $\bar{\delta n^2}$ 
and the shot noise level $\bar{n}$. This ratio is equal to one in the black 
region and increases by a factor of $10^{5/9}\approx 3.6$ for every contour.
The initial increase at low photon numbers $\bar{n}$ is thermal 
($\bar{\delta n^2}\approx \bar{n}^2$).}
\label{noise}
\end{figure}

Since quantum optics textbooks often characterize the light field not by
two time correlations but by the stationary photon number distribution in
the laser cavity, it is interesting to analyze the magnitude of the 
fluctuations given by the variance $\langle \delta\!n^2 \rangle$. 
Using equations (\ref{eq:qcorr}) and (\ref{eq:scalefluct}), an analytical
expression can be derived for the photon number fluctuations. For $\beta\ll 1$
and $\bar{n}\gg n_T$, it reads
\begin{equation}
\label{eq:appfluct}
\frac{\bar{\delta n^2}}{\bar{n}^2} \approx \bigg(1+
      \frac{2\beta \bar{n}^3 \; (2\beta \bar{n} + 1)}
           {(n_T+\frac{1}{2})(4\beta(\kappa\tau+1)\bar{n}^2 +
                  \bar{n}+
                  2\kappa\tau(n_T+\frac{1}{2}))}\bigg)^{-1}.
\end{equation}
Figure \ref{noise} shows a contour plot of the fluctuations as a function
of spontaneous emission factor $\beta$ and photon number $\bar{n}$ for
$3 \kappa\tau=10^4$ and $n_T=3/2$.
It should be noted that the thermal noise region with 
$\delta\!n^2 \approx \bar{n}^2$ extends far beyond the laser threshold
for $\beta > 10^{-6}$. If the
photon number noise threshold $n_\delta$ is defined as the point at which
the photon number fluctuations drop to one half of the thermal
noise level, this threshold is given by 
\begin{equation}
\label{eq:ndeltadef}
      \frac{2\beta n_\delta^3 \; (2\beta n_\delta + 1)}
           {(n_T+\frac{1}{2})(4\beta(\kappa\tau+1)n_\delta^2 +
                  n_\delta+
                  2\kappa\tau(n_T+\frac{1}{2}))}=1.
\end{equation}
This definition of the threshold may be approximated by distinguishing three
types of laser devices, depending on the magnitude of the spontaneous
emission factor $\beta$. The three laser types are
\newcounter{types}
\begin{list}{Type \arabic{case}:}{\usecounter{case}}
\item {\bf Macroscopic lasers}, defined by
$\beta^{-1} > 2 (2\kappa\tau)^2 (n_T+1/2)$, with a noise threshold
$n_\delta=n_{th}$ identical with the laser threshold given by equation 
(\ref{eq:nthdef}).

\item {\bf Mesoscopic lasers}, defined by $4\kappa\tau (n_T+1/2)
<\beta^{-1}< 2 (2\kappa\tau)^2 (n_T+1/2)$, with a noise threshold  
$n_\delta=2\kappa\tau (n_T+1/2)> n_{th}$ slightly above the laser threshold 
given by equation (\ref{eq:nthdef}). 

\item {\bf Microscopic lasers}, defined by 
$\beta^{-1}< 4\kappa\tau (n_T+1/2)$, with a noise threshold $n_\delta
= (\kappa\tau/2) n_{th}$, significantly higher than the laser threshold 
given by equation (\ref{eq:nthdef}). 

\end{list}

Mesoscopic and microscopic lasers therefore have thermal photon number
statistics even above threshold. Recent experimental studies conducted
independently on a solid state laser system seem to confirm this result
\cite{Dru99}.
Note that almost all semiconductor
laser diodes fall into these two categories, since the parameters given 
by equation (\ref{eq:matpar}) indicate that macroscopic semiconductor
lasers must have a spontaneous emission factor $\beta$ smaller than
$10^{-8}$, which corresponds to a threshold current of no less than 50 A.
The borderline between mesoscopic and microscopic semiconductor laser 
devices is found at around $\beta=10^{-4}$ or 5 mA threshold current.
Thus, most modern semiconductor laser devices should exhibit thermal 
photon number fluctuations even above the laser threshold. 

Microscopic lasers show thermal fluctuations even above quantum 
efficiencies greater than 50\% ($2\kappa\bar{n}=j_{th}$). Since high
quantum efficiency is the key to squeezing the laser output by suppressing
the pump noise, such devices can produce a squeezed light output even in the
presence of thermal photon number fluctuations in the laser cavity. 

\section{Low frequency noise}
\label{sec:lfn}

Naturally, it is not possible to measure the light field inside the cavity.
Nevertheless most quantum theories tend to concentrate on the state of the
cavity field modes without any realistic assumptions on the emission process.
It is one of the merits of the original work on squeezing the laser output
\cite{Yam86} that it points out the practical relevance of 
distinguishing between the light inside the cavity and the light emitted by 
the laser device.
On short time-scales, this difference may seem to be irrelevant, because the
emission process only adds some partition noise for times shorter than 
$\kappa^{-1}$ and otherwise produces the same type of statistics as the field
inside the cavity. However, energy conservation introduces a constraint at
longer time-scales. 
Because of this external constraint, the fluctuations of the photon current
outside the cavity may actually be lower than the photon number fluctuations 
inside \cite{Gol97}. If the excitation dynamics are eliminated, this 
constraint affects both the photon number fluctuations inside the cavity
and the low frequency noise equally \cite{Wal94}. If the excitations in the
gain medium provide an additional energy reservoir, however, care must be
taken to distinguish the long term effects of such energy storage on the 
low frequency noise from the short term effects of energy exchange between 
the gain medium and the light field on the photon number fluctuations.

The field outside the cavity represents energy lost from
the laser device, while a time average over the field inside the cavity does 
not have this meaning. In particular, a single photon might stay inside the 
cavity for a long time or for a short time - in the field outside, it will 
only appear once. In order to describe the low frequency part of intensity
fluctuations, it is therefore necessary to discuss the output intensity
$I(t)$ introduced in equation (\ref{eq:intense}). 

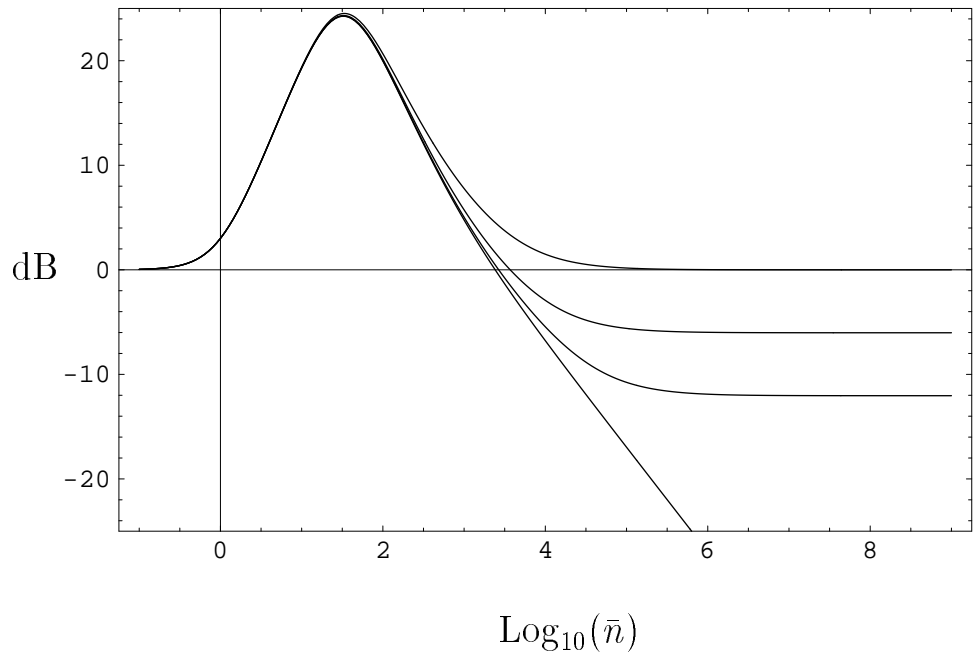
\begin{figure}
\caption{Low frequency noise characteristics in dB relative to the
shot noise limit for $n_T=3/2$ and 
$\beta=10^{-3}$ and pump noise factors of $\sigma=1$, $\sigma=0.25$,
$\sigma=0.0625$ and $\sigma=0$.}
\label{suppression}
\end{figure}

The fluctuations of the average intensity $\bar{I}=2\kappa\bar{n}$ are
given by 
\begin{equation}
\delta I(t) = I(t) - \bar{I} = 2\kappa\delta\!n + q_I(t).
\end{equation}
The two time correlation function of the intensity fluctuation is then 
given by
\begin{equation}
\langle \delta I(t)\delta I(t+\Delta t) \rangle = 
4\kappa^2 \langle \delta\!n(t)\delta\!n(t+\Delta t) \rangle
+ 2\kappa \langle q_I(t)\delta\!n(t+\Delta t) \rangle
+ \langle q_I(t)q_I(t+\Delta t) \rangle.
\end{equation}
The last term represents the shot noise level $L_{\mbox{SN}}$
induced by quantum fluctuations at the cavity mirrors. 
It is therefore convenient to use this term as
a normalization term when performing the time average representing the
limit of low frequencies,
\begin{equation}
\frac{\overline{\delta I^2}(\omega\to 0)}{L_{\mbox{SN}}}
= \frac{\int_0^\infty d\tau \left(
4\kappa^2 \langle \delta\!n(t)\delta\!n(t+\tau) \rangle
+ 2\kappa \langle q_I(t)\delta\!n(t+\tau)\rangle \right)}{\int_0^\infty 
d\tau \langle q_I(t)q_I(t+\tau) \rangle} + 1.
\end{equation}
Using this normalization to the shot noise level, the low frequency limit 
of the intensity noise can be calculated using the linearized Langevin 
equation (\ref{eq:scalefluct}) with the noise terms given by equations
(\ref{eq:qcorr}) and (\ref{eq:output}). For $\beta \ll 1$ and 
$\bar{n}\gg n_T$, the result reads
\begin{equation}
\label{eq:aplfn}
\frac{\overline{\delta I^2}(\omega \to 0)}{L_{SN}}       
      \approx  
     \frac{(n_T+\frac{1}{2})
\left(4\beta \bar{n}^3+2\bar{n}^2+(n_T+\frac{1}{2})\right)}
          {\left(2 \beta\bar{n}^2 + (n_T+\frac{1}{2})\right)^2}
 +\;\sigma\;
     \frac{4\beta^2 \bar{n}^4 +4\beta (n_T+\frac{1}{2}) \bar{n}^3 }
          {\left(2 \beta\bar{n}^2 + (n_T+\frac{1}{2})\right)^2}
.
\end{equation}
Figure \ref{suppression} illustrates the low frequency noise characteristics
for various levels of pump noise suppression $\sigma$. Note that the
low frequency noise below about two times threshold current ($2\kappa\bar{n}
= j_{th}$) does not depend very much on pump noise suppression. Moreover,
the peak value of low frequency noise always coincides with the laser
threshold. Above two times threshold, however, the squeezing obtained
for $\bar{n}\to\infty$ is given by $\sigma$.
In the following, the maximal squeezing potential represented by the 
$\sigma=0$ result will be investigated.

Since squeezing is defined by intensity noise  
below the shot noise limit, the squeezing
threshold $n_{sq}$ for $\sigma=0$ can be defined as the point
at which $\overline{\delta I^2}(\omega \to 0)=L_{SN}$. 
Using $\beta\ll 1$ and $n_T>1/2$, this threshold is found to be given by
\begin{equation}
\label{eq:sqdef}
n_{sq} = \frac{1}{2 \beta} \left((n_T+\frac{1}{2}) + 
\sqrt{(n_T+\frac{1}{2})^2 +(n_T+\frac{1}{2})}\right) \approx 
\frac{n_T+\frac{1}{2}}{\beta} \approx \frac{j_{th}}{2\kappa}.
\end{equation}
The squeezing threshold is thus found at two times threshold current,
corresponding to a quantum efficiency of 50\%. 

\section{Coexistence of squeezing and thermal noise}
\label{sec:sum}

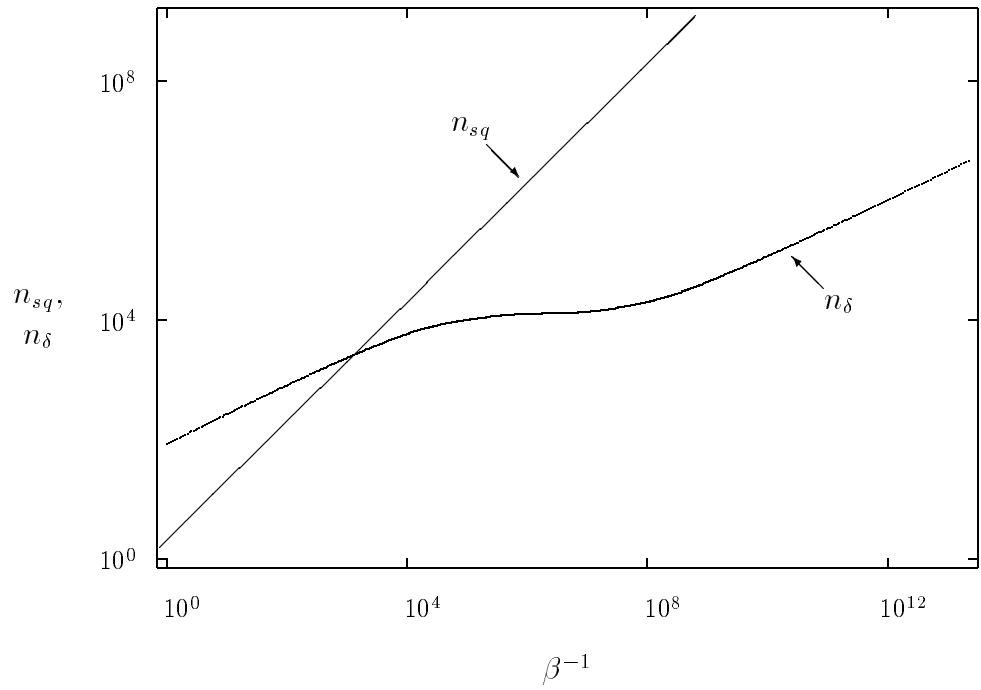
\begin{figure}
\caption{Noise threshold $n_\delta$ and squeezing threshold $n_{sq}$
as a function of the device size given by the inverse spontaneous 
emission factor $\beta^{-1}$. The other device parameters are constant at
$n_T=3/2$ and $3\kappa\tau=10^4$.}
\label{thresholds}
\end{figure}

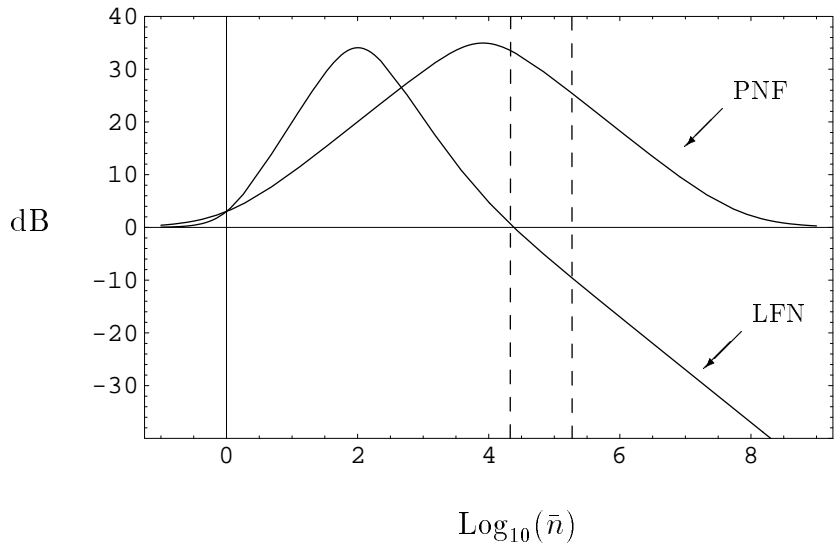
\begin{figure}
\caption{Photon number fluctuations (PNF) and low frequency 
intensity noise (LFN) characteristics in dB relative to shot noise
for $n_T=3/2$ and $3\kappa\tau=10^4$. (a) shows a mesoscopic laser with 
$\beta=10^{-4}$ and (b) shows a microscopic laser with $\beta=10^{-2}$. 
For comparison with the injected current, the dashed lines mark the region 
between twice threshold 
($2\kappa\bar{n}=j_{th}$) and ten times threshold 
($2\kappa\bar{n}=9 j_{th}$).
}
\label{character}
\end{figure}

Above two times threshold ($2\kappa\bar{n}=j_{th}$), the low frequency
noise component of a semiconductor laser device may be squeezed below
the shot noise level by suppressing the noise in the injected current.
All the same, microscopic devices with spontaneous emission factors 
$\beta$ above $10^{-4}$ and corresponding threshold currents below 5~mA
still exhibit thermal photon number fluctuations on short time-scales.
Therefore, an operating regime exists in which squeezing and thermal noise
coexist in the same light field emission. This regime is defined as the 
region between the squeezing threshold given in equation (\ref{eq:sqdef}) and
the noise threshold given in equation (\ref{eq:ndeltadef}). The laser
threshold and the two fluctuation thresholds are shown in 
figure \ref{thresholds}. At about $\beta=10^{-3}$, the two thresholds
cross. Therefore, coexistence of thermal noise and squeezing can be observed
in devices with $\beta>10^{-3}$. Figure \ref{character} shows the photon 
number fluctuations and the low frequency noise for a mesoscopic 
($\beta = 10^{-4}$) and a 
microscopic ($\beta = 10^{-2}$) laser device. 
In the microscopic device, coexistence of thermal noise and squeezing is
clearly observable just above two times threshold ($2\kappa\bar{n}=j_{th}$).

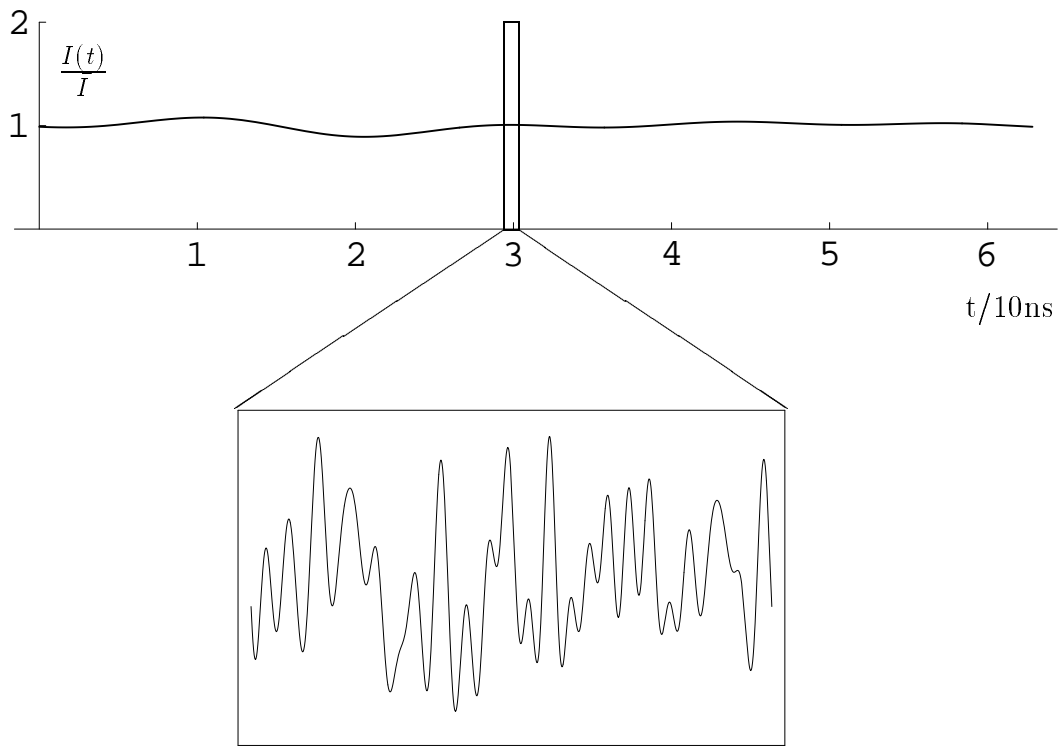
\begin{figure}
\caption{Visualization of the intensity noise. The noise averages out
on long time-scales, even though it is thermal on short time-scales.}
\label{visualnoise}
\end{figure}

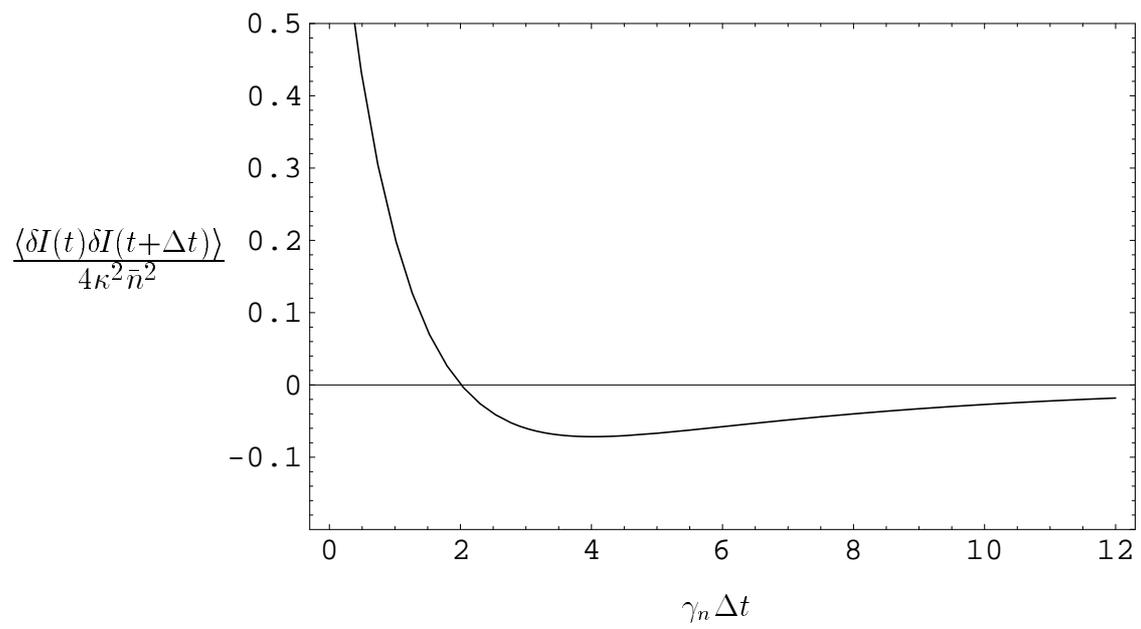
\begin{figure}
\caption{Two time correlation of intensity 
$\langle \delta\!I(t)\delta\!I(t+\Delta t)\rangle$
for $\gamma_n/\Gamma_n=5$.}
\label{twotime}
\end{figure}

What is the relationship between the thermal photon number fluctuations
inside the cavity observed on picosecond time-scales and the suppression
of noise below the shot noise limit on time-scales longer than the 
relaxation rates of the laser dynamics? An attempt to visualize this
relationship is shown in figure \ref{visualnoise}. The short term fluctuations 
average out as the temporal average is taken. This effect is due to the 
anti-correlation of fluctuations at intermediate time differences. In the 
case of over-damped relaxation oscillations, the short term fluctuations 
are given
by a fast optical relaxation $\gamma_n$ and a much slower relaxation of
the carrier system $\Gamma_N$. At a quantum efficiency of 50 \% 
($2\kappa\bar{n}=j_{th}$), the two time correlations of $I(t)$ is 
approximately given by
\begin{equation}
\label{eq:twotime}
\langle \delta\!I(t)\delta\!I(t+\Delta t)\rangle 
\approx 2\kappa\bar{n}\delta(\Delta t)
+ 4 \kappa^2\bar{n}^2 \exp\left(-\gamma_n \Delta t\right)
- 4 \kappa^2\bar{n}^2 \frac{\Gamma_N}{\gamma_n} 
             \exp\left(-\Gamma_N \Delta t\right).
\end{equation}
While the time integrated contribution of the short time bunching is exactly
equal to the time integrated contribution of the long time anti-bunching,
the thermal bunching contributions dominate near $\Delta t=0$ by a ratio
equal to the time-scale ratio $\gamma_n/\Gamma_N \gg 1$. The total low 
frequency noise is then equal to the shot noise term only, because the 
thermal fluctuations 
are anti-correlated on a time-scale of $1/\Gamma_n\approx \tau$ by the slow 
relaxation dynamics of the excitations in the gain medium. 
Figure \ref{twotime} illustrates this transition from bunching to anti-bunching
in the two time correlation of the laser output.

Note that the optical coherence time is equal to or even smaller than
$1/\gamma_n$. In particular, the line-width enhancement effects described
by the $\alpha$ factor and the Peterman factor, respectively, are known 
to cause an additional reduction of the phase coherence not related to 
the photon number relaxation. Therefore, the first order coherence time
will be much shorter than the time during which thermal bunching is observed
in the photon number statistics. This lack of first order coherence suggests
that the squeezing of low frequency intensity noise below the shot noise
limit is only obtained by a summation of the light field intensities of
many modes with independent phase fluctuations. This type of collective
squeezing cannot be attributed to a single coherent light field mode. 
Instead, it should be considered a multi-mode property.

\section{Implications for quantum optics}
\label{sec:col}

The light emitted by a laser propagates in the multi-mode 
continuum of the unconfined electromagnetic field. It is therefore
difficult to describe it in terms of a discrete mode structure. 
As a result, photon number measurements cannot be assigned to 
well defined modes. The randomness of photon detection events is
a consequence of this conceptual difficulty. Nevertheless, time
integrated measurements of photon number can provide precise information
about the number of photons within a given volume. It is tempting to
associate this collective information directly with the single mode
photon number. However, the lack of information available about the
actual photon number distribution among the many modes within the observed
volume should be considered as well. In particular, if all possible
photon number distributions of $n$ photons among $M$ modes are considered
to be equally likely, the situation corresponds to the micro-canonical
ensemble of thermodynamics \cite{LL87}. The density matrix of any mode $i$
which is an arbitrary superposition of the $M$ modes then reads
\begin{equation}
\label{eq:thermal}
\hat{\rho} = \frac{N!(M-1)}{(N+M-1)!}\sum_{n=0}^{N} 
 \frac{(M+N-2-n)!}{(N-n)!} \mid n\rangle\langle n \mid
\;\approx\;  \frac{M}{N+M} \sum_{n=0}^{\infty} \left(1+\frac{M}{N}\right)^{-n}
\mid n\rangle\langle n \mid,
\end{equation}
where the approximation is for large $M$ and $N$.
The photon number distribution of every single mode corresponds to a 
thermal distribution at a temperature proportional to the inverse 
logarithm of $1+M/N$ even though the total 
photon number in the $M$ modes is given precisely by $N$. While the photon
number in each mode fluctuates thermally, the fluctuations in 
the total photon number are suppressed by anti-correlations between
the modes. Such anti-correlations have been described theoretically
\cite{Ken89,Esc96} and observed experimentally between
different longitudinal modes \cite{Ino92,Mar95} and between orthogonal 
polarizations \cite{Kil96} in the squeezed light emission from 
semiconductor lasers. As 
the discussion in this paper indicates, it should be observable in the 
temporal mode structure as well. For the case of over-damped relaxation
oscillations, equation (\ref{eq:twotime}) shows the anti-correlations
between emission modes separated by a time roughly equal to the excitation
lifetime $1/\Gamma_N$. If the coherence length is assumed to be about
$1/\gamma_n$, the intensity distribution given by equation 
(\ref{eq:twotime}) might be interpreted according to equation 
(\ref{eq:thermal}) with $M=\gamma_n/\Gamma_N$ and 
$N=2\kappa\bar{n}/\Gamma_N$. Thus the light field statistics given by
equation (\ref{eq:twotime}) suggest that not a single mode of the total
light field emission is in a squeezed state. If the low frequency noise is
squeezed below the shot noise limit, this effect can be explained as a 
purely collective property without implications for any particular mode. 
Specifically, the slight anti-correlation in photon number between different 
modes shown by the two time correlation function given in equation
(\ref{eq:twotime}) above does not represent quantum mechanical entanglement, 
since there is virtually no phase correlation between the respective modes. 
A semi-classical interpretation of the intensity noise is therefore sufficient
to explain the squeezing properties of the laser field.

\section{Conclusions}
\label{sec:concl}

We have shown that thermal noise in laser cavities may coexist
with squeezing in the low frequency intensity fluctuations. 
The reason for this coexistence is that squeezing is caused by the 
long term anti-correlation of fluctuations caused by the slow loss 
rate of excitations from the gain medium, while the light field 
fluctuations within the optical coherence time are dominated by 
stimulated emission and photon bunching. Consequently,  
squeezing the low frequency intensity fluctuations of a semiconductor
laser diode does not usually reduce the photon number fluctuations
of any single mode. Instead, only the total photon number of a large
number of modes is controlled. This type of collective squeezing should
be distinguished from the single mode squeezing obtained e.g.~by
optical parametric amplification.  
While squeezing in semiconductor laser diodes represents a significant
achievement in controlling the energy flow of light on the quantum level,
it does not produce the entanglement properties which are typically observed
in the single mode squeezing of optical parametric amplifiers.
This limitation severely restricts the use of collectively squeezed light
in quantum optical applications. 

\section*{Acknowledgements}
One of us (HFH) would like to acknowledge support from the Japanese
Society for the Promotion of Science, JSPS.


\vspace{0.5cm}

\begin{thebibliography}{5}
\bibitem{Hak64}
H. Haken, ``Nonlinear Theory of Laser Noise and Coherence I,'' 
Z. Phys. {\bf 181}, 96-124 (1964).

\bibitem{Scu67}
M.O.Scully, W.E.Lamb, ``Quantum Theory of the Optical Maser. I. 
General Theory,'' Phys. Rev. {\bf 159}, 208-226 (1967).

\bibitem{Lax69}
M.Lax and W.H. Louisell, ``Quantum Noise. XII. Density-Operator Treatment of 
Field and Population Fluctuations,'' Phys. Rev {\bf 185}, 568-591 (1969).

\bibitem{Ris70}
H.Risken, ``Statistical Properties of Laser Light,''
p. 239-294 in {\it Progress in Optics}, vol. VIII, 
edited by E.Wolf (North-Holland, Amsterdam 1970).

\bibitem{Hak72}
{\it Laser Handbook}, vol. I,
edited by F.T.Arecchi and O.E. Schulz-Dubois 
(North-Holland, Amsterdam 1972).

\bibitem{Yam86}
Y. Yamamoto, S. Machida, and O. Nilsson, ``Amplitude Squeezing in a
pump-noise-suppressed laser oscillator,'' Phys. Rev. A {\bf 34}, 
4025-4042 (1986).

\bibitem{Mch87}
S. Machida, Y. Yamamoto, and Y. Itaya, ``Observation of Amplitude Squeezing
in a Constant-Current-Driven Semiconductor Laser,'' 
Phys. Rev. Lett. {\bf 58}, 1000-1003 (1987).

\bibitem{Mch88}
S. Machida and Y. Yamamoto, ``Ultrabroadband Amplitude Squeezing in a 
Semiconductor Laser,'' Phys. Rev. Lett. {\bf 60}, 
792-794 (1988).

\bibitem{Mch89}
S. Machida and Y. Yamamoto, ``Observation of Amplitude Squeezing from
Semiconductor Lasers by balanced direct detection with a delay line,'' 
Opt. Lett. {\bf 14}, 1045-1047 (1989).

\bibitem{Ino92}
S. Inoue, H. Ohzu, S. Machida, and Y. Yamamoto, ``Quantum correlation
between longitudinal-mode intensities in a multimode squeezed 
semiconductor laser,'' Phys. Rev. A {\bf 46},
2757-2763 (1992).

\bibitem{Fre93}
M.J. Freeman, D.C. Kilper, D.G. Steel, D. Craig, and D.R. Scifres, 
``Room-temperature amplitude-squeezed light from an injection-locked 
quantum-well laser with a time-varying drive current,''
Opt. Lett. {\bf 20}, 183-185 (1995).

\bibitem{Kil96}
D.C. Kilper, D.G. Steel, D. Craig, and D.R. Scifres, ``Polarization-dependent
noise in photon-number squeezed light generated by quantum-well lasers,'' 
Opt. Lett. {\bf 21}, 1283-1285 (1996).

\bibitem{Ken89}
T.A.B. Kennedy and D.F. Walls, ``Amplitude noise reduction in atomic and
semiconductor lasers,'' Phys. Rev. A {\bf 40}, 6366-6373 (1989).

\bibitem{Wal94}
For a compact explanation of this procedure, see e.g. 
D.F. Walls, G.J. Milburn, {\it Quantum Optics} (Springer, Berlin 1994),
p. 229 ff.

\bibitem{Mar95}
F. Marin, A. Bramati, E. Giacobino, T.-C. Zhang, J.-Ph. Poizat, J.-F. Roch,
and P. Grangier, ``Squeezing and Intermode Correlations in Laser Diodes,''
Phys. Rev. Lett. {\bf 75}, 4606-4609 (1995).

\bibitem{Esc96}
A. Eschmann and C.W. Gardiner, ``Master-equation theory of multimode 
semiconductor lasers,'' Phys. Rev. A {\bf 54}, 760-775 (1996).

\bibitem{Ebe92}
For a practical introduction of the relevant device properties, see e.g.
K.J. Ebeling, {\it Integrated Quantumelectronics}, (Springer, Berlin 1992).
A theoretical derivation from band structure properties can be found in
H.F. Hofmann and O. Hess, ``Quantum Maxwell-Bloch equations for spatially
inhomogeneous semiconductor lasers,'' Phys. Rev. A {\bf 59}, 2342-2358 (1998).

\bibitem{Dru99}
N.J. van Druten, Y.Lien, S.S.R. Oemrawsingh, M.P. van Exter, and J.P. Woerdman,
unpublished.


\bibitem{Gol97}
Y.M. Golubev, I.V. Sokolov, and M.I. Kolobov, ``Possibility of suppressing
quantum light fluctuations when excess photon fluctuations occur inside a
cavity,'' JETP {\bf 84}, 864-874 (1997).

\bibitem{LL87}
L.D. Landau and E.M.Lifschitz, {\it Statistische Physik} (Akademie Verlag,
Berlin 1987), \S 75.

\end{thebibliography}
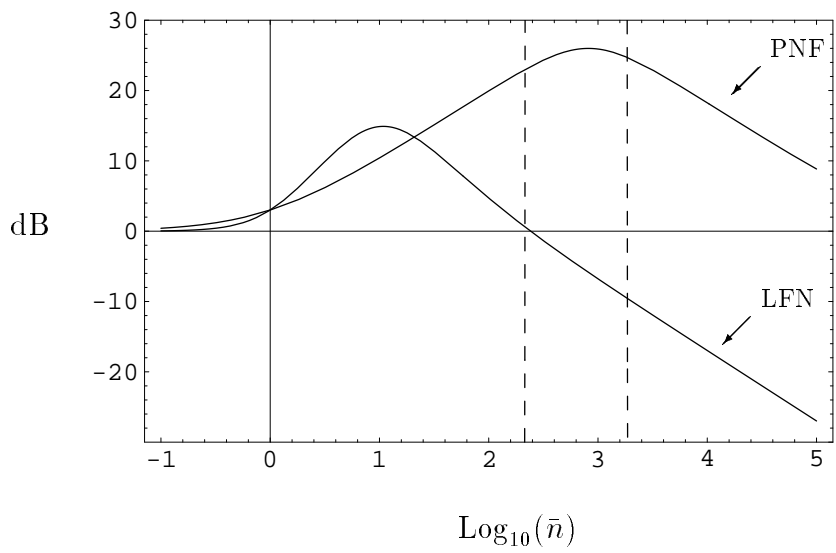
\end{document}